\documentclass{article} 
\usepackage{spconf}  

\usepackage{amsfonts}
\usepackage{amsmath}
\usepackage{graphicx}
\usepackage{url}
\usepackage{eepic, epsfig, amsmath, amssymb, latexsym, setspace, subfig}
\usepackage{rotating}
\usepackage{amsfonts}
\usepackage{amsmath}
\usepackage{graphicx}
\usepackage{nicefrac}
\usepackage{lipsum,multicol}

\newtheorem{theorem}{Theorem}[section]

\numberwithin{equation}{section}

\newcommand{\qed}{\rule{7pt}{7pt}}

\newcommand{\mR}{\mathbb{R}}

\title{MATRIX DESIGN FOR OPTIMAL SENSING}

\name{Hema Kumari Achanta, Weiyu Xu and Soura Dasgupta \thanks{Work  partly supported by US NSF grants CCF-0830747, CNS-1239509 and EPS-1101284, and a grant from the Roy J. Carver Charitable
Trust.}}
\address{Department of ECE, University of Iowa}

\begin{document}
\maketitle

\begin{abstract}
We design optimal $2 \times N$ ($2 <N$)  matrices, with unit columns,  so that  the maximum condition number of all the submatrices comprising 3 columns is minimized. The problem has two applications. When estimating a 2-dimensional signal by using only  three of $N$ observations at a given time, this  minimizes the worst-case  achievable estimation error.  It  also captures the problem of optimum sensor placement for  monitoring a source located in a plane, when only a minimum number of required sensors are active at any given time.   For arbitrary $N\geq3$, we derive the optimal matrices which minimize the maximum condition number of all the submatrices of three columns. Surprisingly, a uniform distribution of the columns is  \emph{not} the optimal design for odd $N\geq 7$.
\end{abstract}


\section{Introduction}
\label{sec:intro}
We consider the problem of designing sensing schemes to optimize the worst-case estimation performance when only a subset of sensors are operational in sensor networks. Consider a set of $N$ sensors which are used to estimate an $M$-dimensional signal, where $N\geq M$.  In our problem, only $K$  out of these $N$ sensors operate at any instant of time. For example, to maximize the lifetime of a sensor network \cite{akyildiz2010wireless, Cardei:2005:IWS:1160086.1160098,Petropulu, Pramod, Oshman, Savkin}, at any single time instant, only $K$ sensors are turned on to monitor the $M$-dimensional signal. If we assume that each time these $K$ sensors are uniformly selected from the $\binom{N}{K}$ possible subsets, on average the lifetime of the sensor network is extended by a factor of $N/K$.  As another example, in hostile environments such as battlefields, it is very common that only a limited number of sensors, say $K$ out of $N$,  are able to survive and operate as designed. In these scenarios, while we
only have a limited sensing resources at a single time instant, we wish to achieve the best estimation  from limited observations. It is thus useful to maximize the worst-case performance of the sensing system, no matter what set of sensors are used or  survive. We thus study the design of   sensing schemes that optimize worst-case performance. Before a formal mathematical formulation, we review two sensor network applications which relate singular values of certain matrices to estimation performance.

\subsection{Signal Estimation} \label{ssens}
With $x\in \mR^M$ representing the  signal, consider a sensing matrix $A \in \mR^{M\times N}$. Each of the  $N$ sensors generates a real observation represented by an    inner product between  $x$ and a column of $A$. Let ${KS} \subseteq \{1,2..., N\}$, with cardinality $|KS|=K$, be the subset of sensors that are active at a given time. The measurement matrix of the active sensors is then $A_{{KS}}  \in \mR^{M\times K}$  consisting of the $K$ columns of $A$ indexed by $KS$.
With noise $w$, the  measurement  $y \in \mR^K$ is
\begin{equation}\label{eq:signal}
y=A_{KS}^{T}x+w,
\end{equation}

Suppose the  singular values of $A_{KS}$ are $\sigma_i$. Then as long as $A_{KS}$ has full row rank, the  estimation  error  satisfies
\begin{equation*}
\|\hat{x}-x\|_2=\|(A_{KS}A_{KS}^{T})^{-1}A_{KS}(w)\|_2 \leq \frac{\|w\|_2}{\sigma_{min}}.
\end{equation*}
To optimize the worst-case performance, we must design  $A$ to maximize the smallest singular value among all the $\binom{N}{K}$ possible submatrices $A_{KS}$. To make the problem meaningful, we assume that each column of $A$ has unit $\ell_2$ norm.  When $M=2$, this is equivalent to minimizing the maximum condition number among all $\binom{N}{K}$ submatrices $A_{KS}$.

\subsection{Source Monitoring in the Plane}
A second motivating application for this paper is optimum sensor placement for source monitoring in  $\mR^2$,   \cite{DID2}-\cite{MB}.
Monitoring is related to the notion of localization, where several sensors collaborate to locate a source, using some relative position information. The latter could be distance, bearing, time of arrival, time difference of arrival
or received signal strength (RSS). Monitoring assumes that a hazardous source has already been located at some $z\in \mR^2$, and a group of sensors at
$x_i \in \mR^2$ monitor it by continuously estimating its position from a safe distance. Thus \cite{DID2}-\cite{MB}  place  sensors, i.e.  choose $x_i$, so that  the minimum eigenvalue of the Fisher Information Matrix (FIM)  underlying the estimation problem is maximized. This ensures that under continuous monitoring and
Maximum Likelihood (ML) estimation, asymptotically, the mean-square  error in estimating $z$, is minimized, \cite{coverIT, Scharf1993301, van2001detection}.  As $z$ is at least roughly known, so also is the FIM.

Consider \cite{DID2,Ibeawuchi.S}, where
no sensor can be closer than $D$ from the source. Each  measures the RSS of the signal emanating from the source under  log-normal shadowing, i.e.  with known positive real scalars $A$ and $\beta$, the RSS $s_i$ at the $i$-th sensor obeys, for mutually independent $w_i\sim {\cal N} \left( 0,\sigma^2\right )$:
\begin{equation} \label{RSS}
\ln s_{i} = \ln A - \beta \ln ||x_{i}-z||+w_{i},
\end{equation}
The underlying FIM with $N$-sensors is, \cite{DID2}
\begin{equation} \label{FIM1}
G= \frac{\beta^2}{\sigma^2(\ln 10)^2}\sum_{i \in N}\frac{(x_{i}-z)(x_{i}-z)^{T}}{||x_{i}-z||^4}.
\end{equation}
The optimal sensor placement problem then becomes: Given, $z\in \mR^2$, and $D>0$, find $x_i \in \mR^2$, $i\in \{1,\cdots, N\}$ so that the minimum eigenvalue of $G$ is maximized, subject to:
$
||x_{i}-z||\geq D.
$
Because of the denominator in (\ref{FIM1}), the minimum eigenvalue of $G$ is maximized only if for all $i\in \{1,\cdots, N\}$, $\|x_i-z\|=D$. Without loss of generality one can assume $D=1$ and $z=0$. Thus  effectively one must maximize the minimum eigenvalue of
\begin{equation} \label{FIM12}
 \sum_{i \in N}x_ix_i^T,
\end{equation}
subject to $\|x_i\|=1$. This is tantamount to minimizing the condition number of $F$ as its trace is constrained to be $N$.

Now suppose to prolong battery life, only a subset of sensors is activated at a given time,  \cite{Oshman,Savkin}. The logical problem to consider is then for some $K$, $KS$ as defined above, and
\begin{equation} \label{FIM}
F_{KS}= \sum_{i \in KS}x_ix_i^T,
\end{equation}
to minimize the largest condition number of $F_{KS}$, among all $KS\subseteq \{1,\cdots, N\}$. With $A_{KS}$ having columns $x_i$, $i\in  KS$, we have $F_{KS}=A_{KS}A_{KS}^T$, and a similar setting of Section \ref{ssens}. We observe, that the {\it minimum $K$ needed for source monitoring is three, motivating the rest of this paper where $K=3$ is considered.} In particular RSS provides a distance estimate. Distances from three non-collinear sources are necessary to localize, \cite{FDA}. This scenario also applies to
the case where only three sensors survive hostilities.

The rest of this paper is organized as follows. Section \ref{sec:formulation} gives a precise mathematical formulation. Section \ref{sec:formula} provides a formula for the minimum condition number of submatrices when $M=2$. Section \ref{sec:Neven} characterizes optimal solutions all for $M=2$, $K=3$ and arbitrary $N\geq 3$. Section \ref{sec:Sim} presents  simulations.

\section{Problem Formulation}
\label{sec:formulation}

  Let $M \leq N$ be positive integers and $A=[a_{1}, a_{2}, ...,  a_{N}]$, where $a_{i}\in \mR^M$ obey  $||a_{i}||_2=1$ for $1\leq i \leq N$. Let ${KS} \subseteq \{1,2,...,N\}$ be a subset with cardinality $|KS|=K$. Now, $A_{{KS}} \in \mR^{M \times K}$ is the submatrix $A_{{KS}}=[a_{i_{1}}, a_{i_{2}},.....,a_{i_{K}}]$ with columns indices $i_j$, $1\leq j \leq K$, from the set $KS$. 
Then our optimal design problem for the parameter set $(M,N, K)$ is:
\begin{equation*}
\max_{A \in \mR^{M \times N} \text{with unit-normed columns}}  \left\{ \underset{{KS\subseteq \{1,2,...,N\}}}{\text{min}} {\sigma_{min}(A_{KS})}\right\}.
\end{equation*}
For $M=2$, this is equivalent to minimizing the condition number:
\begin{equation*}
\min_{A \in \mR^{M \times N} \text{with unit-normed columns}}  \left\{ \underset{{KS\subseteq \{1,2,...,N\}}}{\text{max}} \frac{\sigma_{max}(A_{KS})}{\sigma_{min}(A_{KS})}\right\}.
\end{equation*}

Note the similarity between this problem and the problem of designing compressive sensing matrices \cite{CS} satisfying the restricted isometry property (RIP), which also requires the condition numbers for the submatrices be small. As opposed to the design of compressive sensing matrices satisfying RIP \cite{CS}, in our problem, the submatrices $A_{KS}$ are wide  rather than tall. The motivating applications are also different from compressive sensing.

As noted earlier, motivated in part by  2-dimensional source monitoring with the minimum number of sensors i.e. $K=3$, we restrict attention to the case of $K=3$ and $M=2$, where closed form expressions are possible and surprising conclusions, that may illuminate the problem solution for higher values of $K$ and $M$, are obtained.

\section{Derivation of the Condition Number for $M=2$}
\label{sec:formula}


The condition number of $\tilde{A}_{KS}=A_{KS}A_{KS}^T$ is given by
\begin{equation}
\kappa(\tilde{A}_{KS}) =\frac{\max_{||\eta||=1}(\eta^{T}\tilde{A}_{KS}\eta)}{\min_{||\eta||=1}(\eta^{T}\tilde{A}_{KS}\eta)}
\end{equation}
Since the columns of $A$ are unit-normed, we can represent $A=[a_1, a_2, ...., a_N]$ with
\begin{equation}
a_{i}=\left( \begin{array}{cc}
\cos \theta_{i} &
\sin \theta_{i}\end{array} \right)^T
\end{equation}
for $1\leq i \leq N$, where $\theta_{i}\in [0,\pi)$ (we do notice shifting $\theta_i$ by $\pi$ will not change the condition number of any submatrix). Since $||\eta||_2=1$  we can choose
$
\eta =\left( \begin{array}{cc}
\cos \alpha &
\sin \alpha\end{array} \right)^T.$
Thus
\begin{equation*}
 \eta^{T} \tilde{A}_{KS} \eta = \frac{{K}}{2}+ \frac{1}{2}\sum_{j=1}^{{K}}\cos(2(\alpha-\theta_{i_j})) =J(\alpha).
\end{equation*}
Let us define
\begin{equation}
 J(\alpha) =\frac{{K}}{2}+ \frac{1}{2}\sum_{j=1}^{{K}}\cos(2(\alpha-\theta_{i_j})).
 \label{eq:Jdefinition}
\end{equation}
Then the minimum  (maximum) eigenvalue of $\tilde{A}_{KS}$ is achieved when $J'(\alpha)=0$ and $J''(\alpha) > (<)0$.
With
\[
\gamma={(\sum_{j=1}^{{K}}\sin(2\theta_{i_j}))^2+(\sum_{j=1}^{{K}}\cos(2\theta_{i_j}))^2} \neq 0,
\]at a minimum or maximum, $\alpha$ satisfies
\begin{equation*}
 \cos(2\alpha)= \sum_{j=1}^{{K}}\cos(2\theta_{i_j})/\gamma
\end{equation*}
and
\begin{equation*}
  \sin(2\alpha)= \sum_{j=1}^{{K}}\sin(2\theta_{i_j})/\gamma.
\end{equation*}
Thus,
\begin{equation}
 J(\alpha) =\frac{{K}}{2}+ \frac{1}{2}\sum_{j=1}^{{K}}\cos(2\alpha)\cos(2\theta_{i_j})+\frac{1}{2}\sum_{j=1}^{{K}}\sin(2\alpha)\sin(2\theta_{i_j}).
 \label{eq:Jexpansion}
\end{equation}
Combining the optimizing $\alpha$ and (\ref{eq:Jexpansion}), we have
\begin{eqnarray}
&&J(\alpha) = \frac{{K}}{2} +\nonumber\\
&&\frac{1}{2}
\frac{\sum_{j=1}^{{K}}\sum_{l=1}^{{K}}(\cos(2\theta_{i_l})\cos(2\theta_{i_j})+ \sin(2\theta_{i_l})\sin(2\theta_{i_j})) }
{\sqrt{(\sum_{l=1}^{{K}}\sin(2(\theta_{i_l})))^2+(\sum_{l=1}^{{K}}\cos(2\theta_{i_l}))^2}}.\nonumber
\label{eq:Jexpansion2}
\end{eqnarray}

On simplification, the maximum and minimum eigenvalues of $\tilde{A}_{KS}$ are given by
\begin{equation}
J(\alpha_{max})=\frac{{K}}{2}+ \frac{1}{2}\sqrt{\frac{{K}}{2}+\sum_{j=1}^{{K}}\sum_{l=j+1}^{{K}} \cos2(\theta_{i_l}-\theta_{i_j})},
\label{eq:Jmax}
\end{equation}
and \begin{equation}
J(\alpha_{min})=\frac{{K}}{2}- \frac{1}{2}\sqrt{\frac{{K}}{2}+\sum_{j=1}^{{K}}\sum_{l=j+1}^{{K}} \cos2(\theta_{i_l}-\theta_{i_j})}.
\label{eq:Jmin}
\end{equation}
respectively. Thus minimizing the condition number of $\tilde{A}_{{KS}}$ for a given set of indices $\{i_{1},i_{2},..,i_{{K}}\}$ is the same as (the equation inside the square root is always nonnegative)
 \begin{equation}
\begin{aligned}
& &\min_{{\theta_{i_1}, ..., \theta_{i_K}}}
& &  {\sum_{j=1}^{{K}}\sum_{l=j+1}^{{K}} \cos2(\theta_{i_l}-\theta_{i_j})}.
\end{aligned}
\label{eq:doubling}
\end{equation}

With $KS\subseteq \{1,2,...,N\}$,  the optimal sensing matrix design problem for $M=2$ can be reformulated as,
\begin{eqnarray*}
\min_{\theta_{1}, ..., \theta_{N}}  \underset{KS=\{i_{1},i_{2},..,i_{K}\} } {\text{max}}
{\sum_{j=1}^{{K}}\sum_{l=j+1}^{{K}} \cos2(\theta_{i_l}-\theta_{i_j})}.\\
\end{eqnarray*}
%

In the following sections, we will derive the optimal design for $K=3$, which has important applications in location monitoring in sensor networks.

\section{Optimal placement}\label{sec:Neven}

We now consider solutions  for $M=2$, $K=3$ and different values of $N$.
\subsection{$K=3$, $N$ is an even number}

\begin{figure}[htbp]
\begin{center}
\includegraphics[height=3in, width=3in]{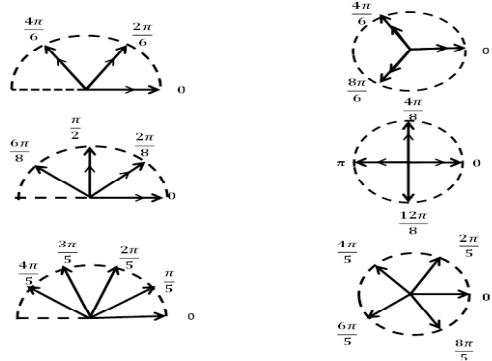}
\vspace{-15mm}
\caption{Illustration of angle arrangements $\theta_i$'s for $N=6$, $7$ and $5$ respectively, using the $3$ rows of figures from top to bottom. The left figures represent the angle ($\theta_i$) for the columns of sensing matrices. Right figures are doubling those angles ($2\theta_i$) as in the objective function in (\ref{eq:doubling}). } \label{monitoring9} 
\end{center}
\end{figure}

 For even-numbered $N$, the optimal design is given as below.
\begin{theorem}
If $K=3$ and $N$ is an even number, then the set of angles (a)  $\theta_i=\frac{2\pi (i-1)}{N} \mod \pi$, $1 \leq i \leq  N$, or (b) $\theta_i=\frac{2\pi (i-1)}{N} $, $1 \leq i \leq  N$, minimizes the maximum condition number among all sub-matrices with $K$ columns.
\label{thm:K3NEven}
\end{theorem}
Observe, (a) actually aligns pairs of angles together (see Fig \ref{monitoring9}) and is not useful for source monitoring where at least three distinct sensor locations are necessary, \cite{FDA}. On the other hand (b) leads to distinct locations by separating adjacent sensors $2\pi/N$ radians apart.

\subsection{$K=3$, $N=3$ or $5$}
\label{sec:N5}
These stand apart from other odd $N$ values:
\begin{theorem}
Let $K=3$ and $N=3$ or $5$. Then the set of angles $\theta_i=\frac{\pi (i-1)}{N}$, $1 \leq i \leq  N$, minimizes the maximum condition number among all sub-matrices with $K=3$ columns.
\label{thm:K3N5}
\end{theorem}

\subsection{$K=3$, $N\geq7$ is an Odd Number}
\label{sec:Ngeq7}
One might think that the uniform distributed design is optimal for $N\geq 7$. However, this is not true from the following theorem. Instead, the optimal design is to eliminate one angle from the optimal design for $(N+1)$.
\begin{theorem}
If $K=3$ and $N \geq 7$ is an odd number, then $\theta_i=\frac{2\pi (i-1)}{N+1} \mod \pi$, $1 \leq i \leq  N$, minimizes the maximum condition number among all sub-matrices with $K=3$ columns.
\label{thm:K3Ngeq7}
\end{theorem}

\section{Simulation Results}
\label{sec:Sim}
We now present  simulation results.
\subsection{Worst Case Condition Number vs $N$}
\begin{figure}[!htb]
\includegraphics[height=2 in, width=3.5in]{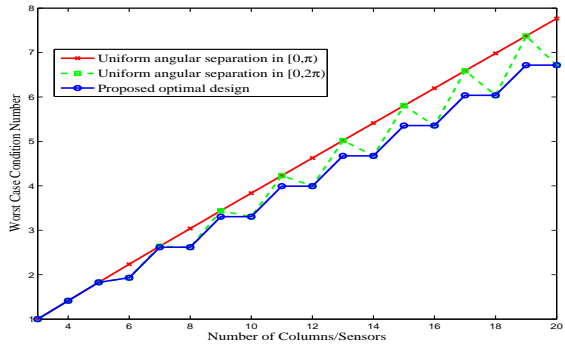}
\caption{Worst case condition number versus $N$} \label{fig:}
\end{figure}
We compare the maximum condition number among all the possible $2 \times 3$ submatrices in  three different cases shown in Fig.\ref{fig:}. The cases are, (i) when successive sensors are placed in a semicircle $\pi/N$ apart, namely $\theta_i=0,\frac{\pi}{N},...,\frac{\pi (N-1)}{N}$, (ii) they are placed $2\pi/N$ apart, namely $\theta_i=0,\frac{2\pi}{N},...,\frac{2\pi (N-1)}{N}$, and (iii) they are placed in a manner specified by our theorems. That the performance of  (ii) matches (iii) for even $N$ conforms with earlier observations.

\subsection{Worst Mean Square Signal Estimation Error vs $N$}
\begin{figure}[!htb]

\includegraphics[height=2 in, width=3.5in]{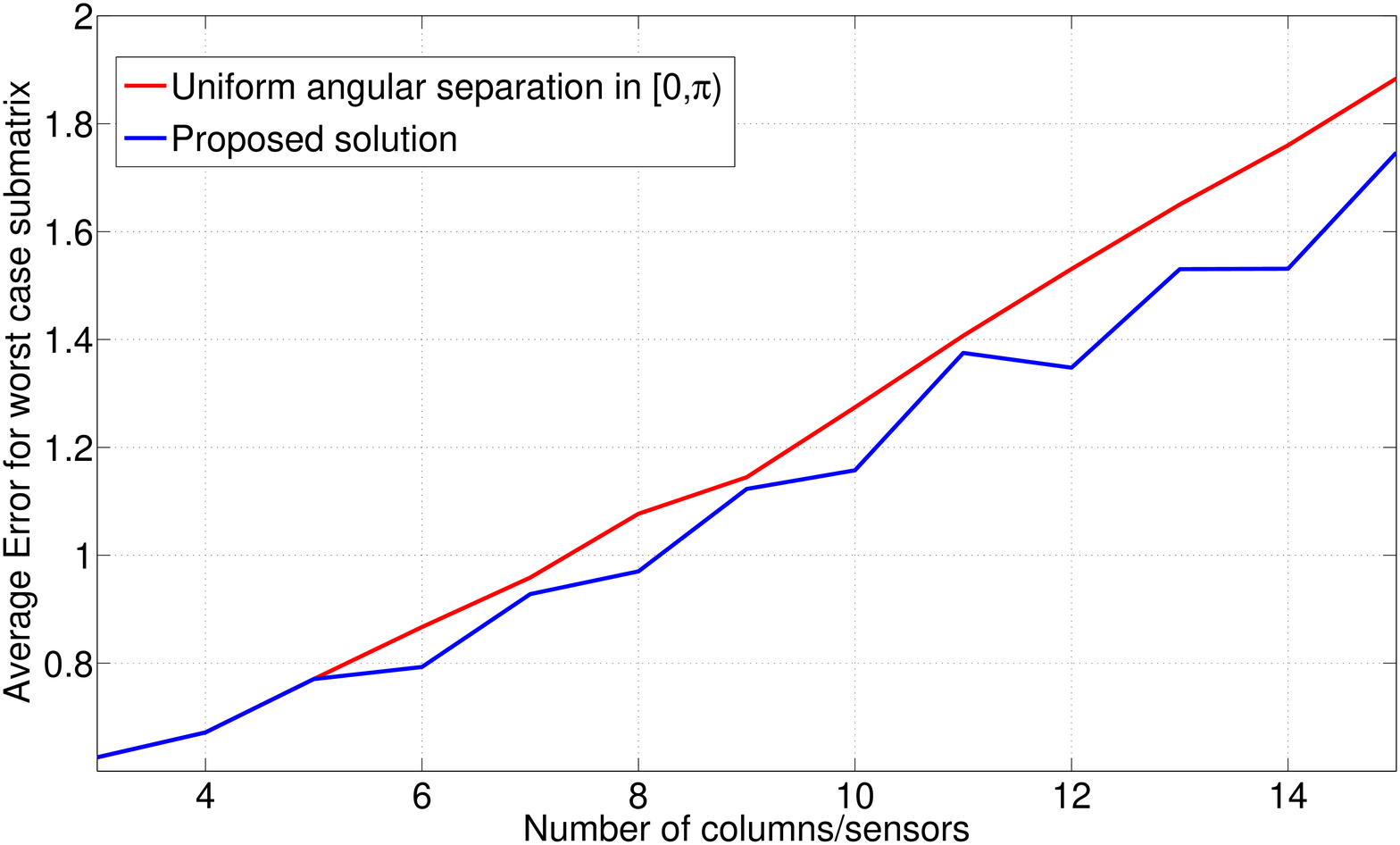}
\caption{Worst case estimation error versus the number of columns in the sensing matrix.}
\label{fig:worsterror}
\end{figure}

Consider the setting of Section \ref{ssens}.
We compare in Fig. \ref{fig:worsterror} the mean square error (MSE) for worst-case submatrices yielded by  (i)  above with that yielded by the postulated optimum  for  sensors ranging in number from 3 to 15. The signal $x$ in (\ref{eq:signal}) is $[9,9]^T$. The noise in each measurement is  ${\cal N} \sim (0,1)$. For each value $N$, the estimation error $||\hat{x}-x||^2$ for worst-case submatrices was averaged over 2000 instances. Again the predicted optimal placement is superior.

\subsection{Monitoring Error vs SNR}
\begin{figure}[!htb]
\includegraphics[height=2 in, width=3.5in]{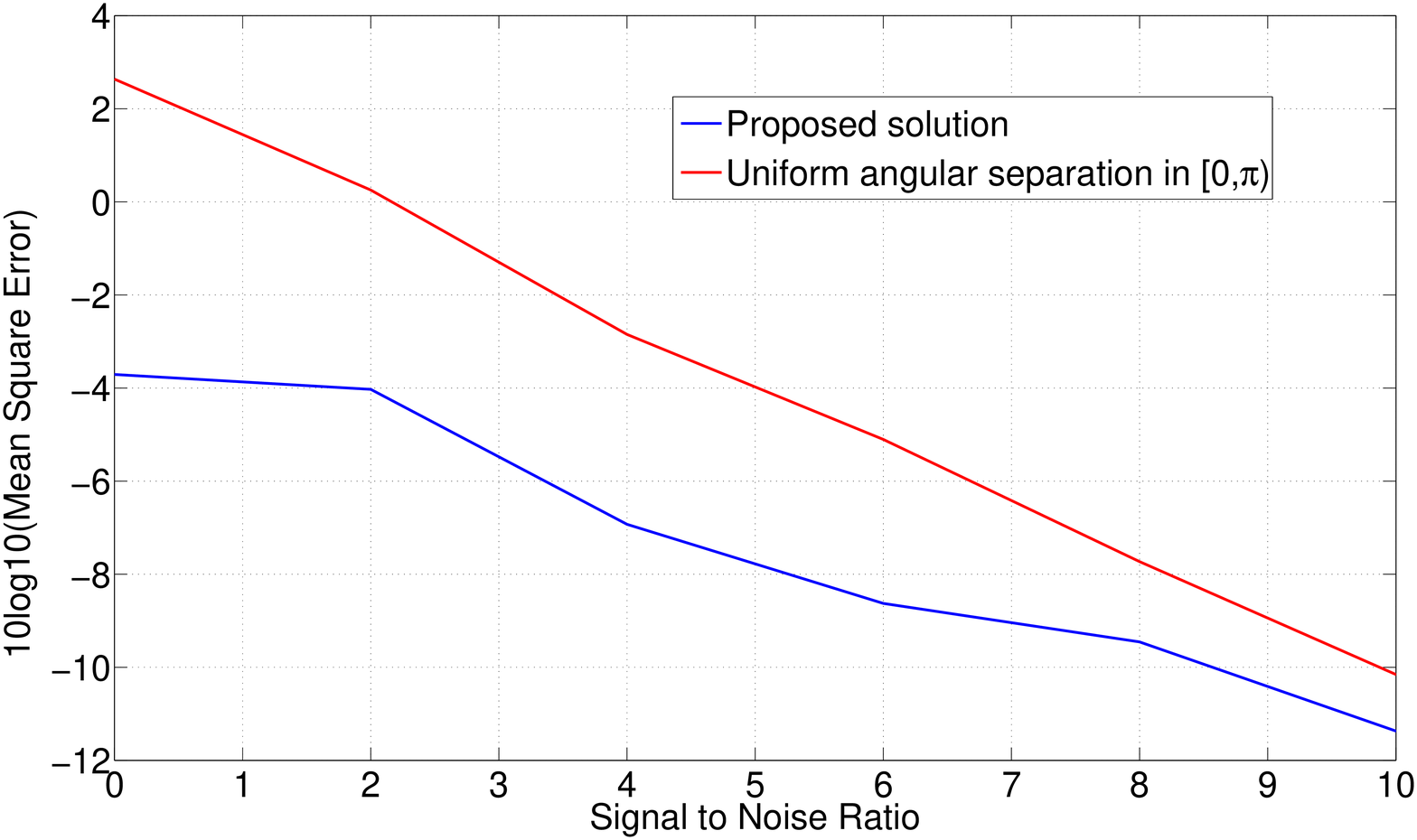}
\caption{Mean square error(dB) in the source location estimate when the worst performing subset of sensors are active versus the Signal to Noise Ratio(dB).} \label{errorvsmse9}
\end{figure}
Fig. \ref{errorvsmse9} compares the ML estimation of a source at the origin with $N=10$, from RSS under log-normal shadowing in the case where the sensors are placed as in (i) against optimal placement. The latter's superiority is evident.

\section{Conclusion and Future Work}
\label{sec:conclusion}
We propose the problem designing optimal $M \times N$ ($M \leq  N$) sensing matrices which minimize the maximum condition number of all the submatrices of $K$ columns.  Such matrices minimize the worst-case estimation errors when only $K$ sensors out of $N$ sensors are available for sensing at a given time. When $M=2$ and $K=3$, for an arbitrary $N\geq3$, we derive the optimal matrices which minimize the maximum condition number of all the submatrices of $K$ columns. It is interesting that minimizing the maximum coherence between columns does not always guarantee minimizing the maximum condition number.
\bibliographystyle{plain}

\end{document}